\documentclass[aps,prl,twocolumn,groupedaddress]{revtex4-1}
\usepackage{graphicx,color}
\usepackage{dcolumn}
\usepackage{amsmath}

\usepackage{float}
\hfuzz=\maxdimen
\begin{document}
\title{ The out-of-plane magnetoresistance in a Van der Waals thin film of WTe$_2$}
\author{Y.~S.~Liu$^{1,2,3\star}$}
\author{H.~Xiao$^{4}$}
\author{C.~Zhang$^{5}$}
\author{C.~W.~Zhang$^{6}$}
\author{Y.~G.~Shi$^{6}$}
\author{T.~Hu$^{3}$}
\author{C.~M.~Schneider$^{1,2\dag}$}
\affiliation{$^{1}$Peter Gr{\"u}nberg Institute PGI-6, Forschungszentrum J{\"u}lich, D-52425 J{\"u}lich, Germany}
\affiliation{$^{2}$Fakult{\"a}t f{\"u}r Physik, Universit{\"a}t Duisburg-Essen, D-47057 Duisburg, Germany}
\affiliation{$^{3}$Beijing Academy of Quantum Information Sciences, Beijing, 100193, China}
\affiliation{$^{4}$Center for High Pressure Science and Technology Advanced Research, Beijing, 100094, China}
\affiliation{$^{5}$State Key Laboratory of Functional Materials for Informatics, Shanghai Institute of Microsystem and Information Technology, Chinese Academy of Sciences, Shanghai 200050, China}
\affiliation{$^{6}$Institute of Physics, Chinese Academy of Sciences, Beijing 100190, China}

\begin{abstract}
We report the magneto-transport measurements of thin film devices of the topological Weyl semimetal WTe$_2$ with the applied current along and vertical to the in-plane directions. The device is composed of a Van der Waals thin film of WTe$_2$ sandwiched between top and bottom Au electrodes. At low temperatures, we found a large unsaturated in-plane magnetoresistance and a saturated out-of-plane magnetoresistance when the external magnetic fields are applied perpendicular to the plane. By analysis of Shubnikov-de Haas oscillations, one  oscillation peak is found in the out-of-plane magnetoresistance, in contrast to four oscillation peaks  in the in-plane magnetoresistance. Our work provides new insight into the origin of the unsaturated magnetoresistance in WTe$_2$ and may inspire non-planar engineering to reach higher integration in spintronics.
\end{abstract}


\maketitle

\section{introduction}
WTe$_2$ is a layered transition-metal dichalcogenide (TMD) consisting of a tungsten layer surrounded by two tellurium layers stacked along the $z$ axis.  It  exhibits an extremely large  magnetoresistance (XMR) and shows no sign of saturation in high fields.\cite{ali2014large} The XMR in the type-II Weyl semimetal WTe$_2$ recently stimulates extensive theoretical and experimental magneto-transport investigations.\cite{cai2015drastic,zhu2015quantum,thoutam2015temperature,li2017evidence,pletikosic2014electronic}
ARPES experiment  in WTe$_2$ suggests that the XMR is attributed to the compensation between the balanced electron and hole populations,\cite{pletikosic2014electronic} like the  behaviour expected in a perfectly-compensated semimetal.\cite{pippard1989magnetoresistance} In contrast to the normal semimetal, WTe$_2$ is a type-II Weyl semimeta, where Fermi surfaces consist of a pair of electron- and hole- pockets contacting at the Weyl node.\cite{soluyanov2015type,li2017evidence}  Shubnikov-de Haas effect (SdH) oscillations, Seebeck and Nernst measurements support the Fermi surface consisting of two pairs of electron-like and hole-like pockets\cite{zhu2015quantum,cai2015drastic}.

With the  reduction of  thickness, a metal-to-insulator transition is observed in thin WTe$_2$ flakes.\cite{wang2015tuning}  Gate-tunable  magnetoresistance is found in ultra-thin  WTe$_2$.\cite{wang2016gate,liu2017gate} In atomically thin WTe$_2$, the magnetoresistance can be tuned from positive to negative.\cite{zhang2017tunable} In addition, by electrostatically doping ultra-thin WTe$_2$, the XMR is turned on and off.\cite{fatemi2017magnetoresistance}
In a monolayer crystal WTe$_2$,  the quantum spin Hall effect\cite{wu2018observation} and Landau quantization\cite{wang2021landau}  are observed. In bilayers of WTe$_2$, the nonlinear Hall effect is observed in spite of its non-magnetic and time-reversal-symmetric conditions.\cite{ma2019observation} It needs further investigation in the WTe$_2$ ultrathin flake, where the electronic structure is modulated by the dimension.\cite{xiang2018thickness}

In this paper, we fabricate a thin film device of WTe$_2$ with top and bottom  electrode structure. By choosing the current directions, we performed in-plane  and  out-of-plane magneto-transport measurements in this device. We found a large unsaturated in-plane magnetoresistance and a saturated out-of-plane magnetoresistance. The  Shubnikov-de Haas oscillations are also observed in the  magnetoresistance at both directions at low temperature. By analysis of the oscillations, we found four oscillation peaks  in the in-plane magnetoresistance, but  one  oscillation peak  in the out-of-plane magnetoresistance. 

\section{Experimental details} \label{exp}
\begin{figure*}
	\includegraphics[trim=0cm 0cm 0cm 0cm, clip=true, width=0.9\textwidth]{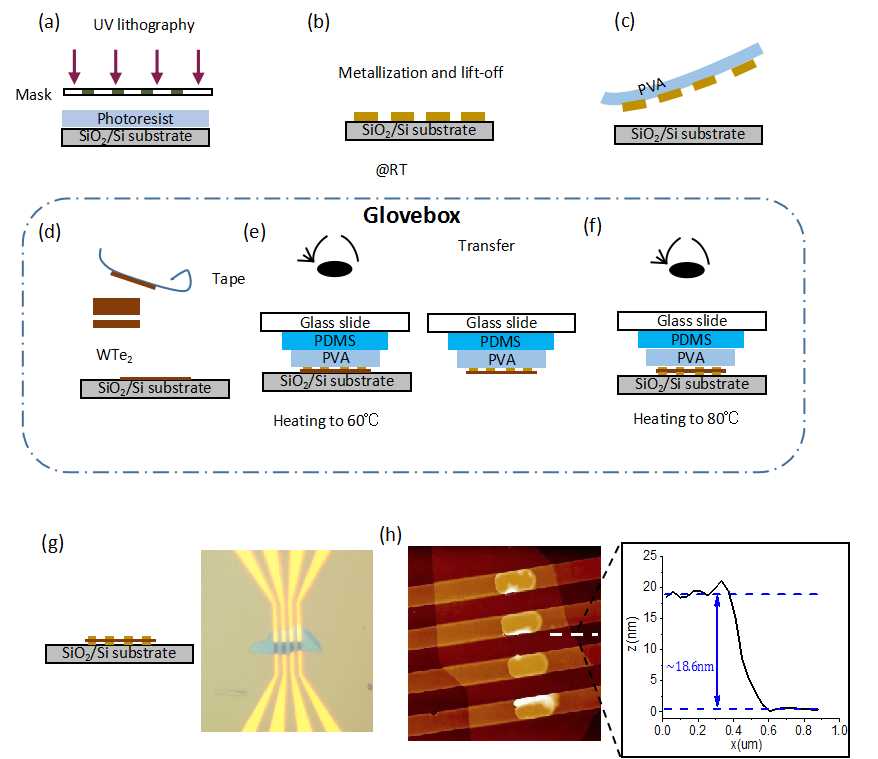}
		\caption{Schemetics of the fabrication process.
			(a) UV lithography patterning onto the SiO$_2$/Si substrate.
			(b) Au are deposited onto the SiO$_2$/Si substrate and Au electrodes are left after a standard lift-off process applied to remove the photoresist.
			(c)	The Au electrodes are peeled off by polyvinyl alcohol (PVA) at room temperature from the SiO$_2$/Si substrate.
			(d) Preparation of exfoliated WTe$_2$ flake on SiO$_2$/Si substrate in glovebox.
			(e) The Au/PVA/PDMS prepared on glass slide are in contact with the thin WTe$_2$ film when heating the stage to 60$^o$C. After heated for 2 minutes, the WTe$_2$/Au/PVA/PDMS structure is obtained after gentle separation of the glass slide from the SiO$_2$/Si substrate.
			(f) Adjusting the relative position of Au electrodes on SiO$_2$/Si substrate and WTe$_2$/Au/PVA/PDMS on glass slide under an optical microscope and make  gentle contact. The stage is kept heating to 80$^o$C  for 2 minutes.
			(g) After gentle separation from the PDMS on the glass slide, the  Au-electrodes/WTe$_2$/Au-electrodes device is obtained. Schematic structure and optical image of the device are shown.
			(h) AFM micrograph of fabricated WTe$_2$ vertical structures show the vertical channel region. Height profile (right inset) is extracted along the dotted line.
}
	\label{fig:Figure 1}
\end{figure*}

\begin{figure}
	\includegraphics[trim=0cm 0cm 0cm 0cm, clip=true, width=0.45\textwidth]{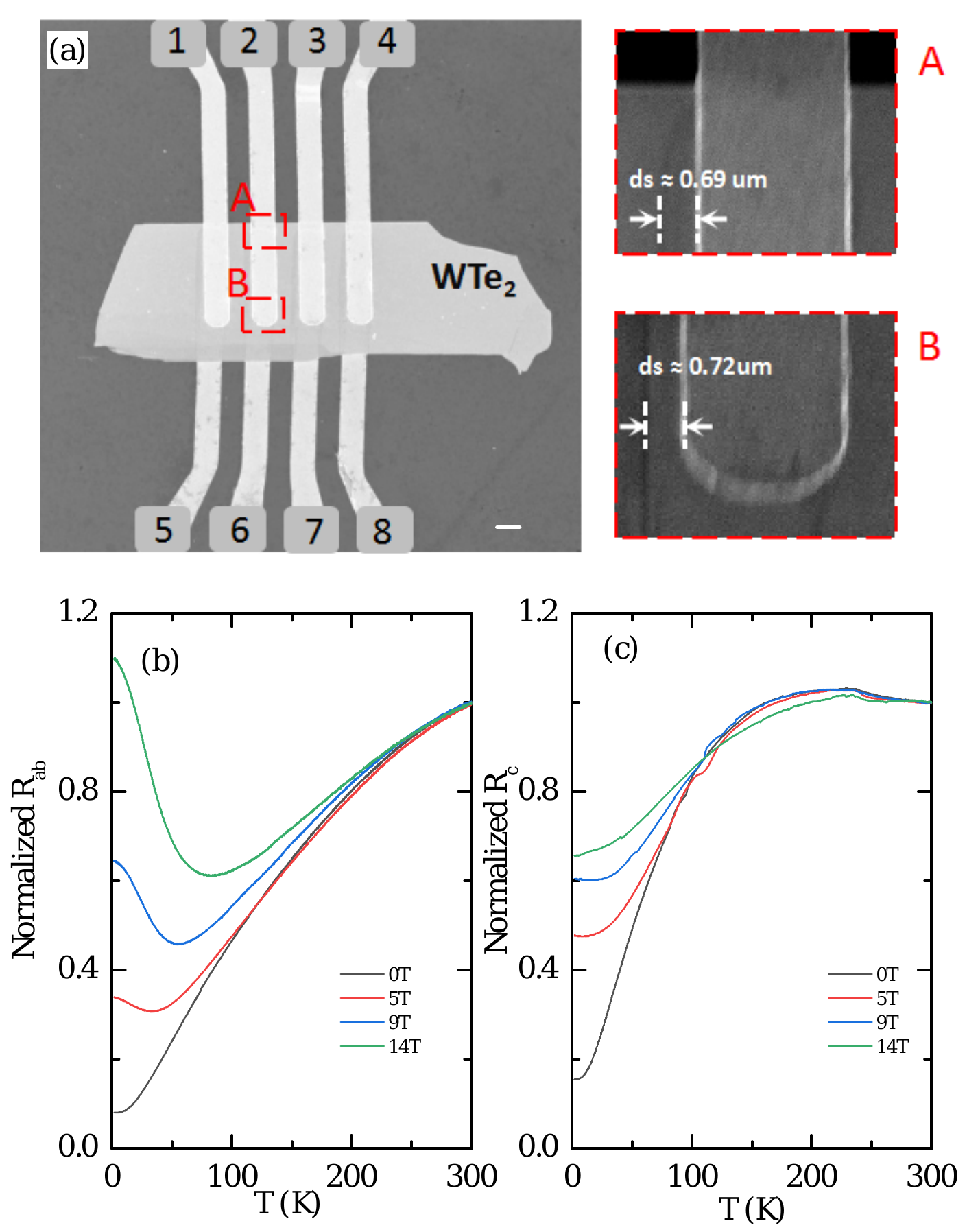}
\caption{ Structure and transport behavior of the WTe$_2$ device. 	(a) Scanning electron micrographs of the  WTe$_2$ device, consisting of the ultrathin WTe$_2$ film sandwiched between the top (marked with 1, 2, 3 and 4) and bottom (marked with 5, 6, 7 and 8) gold electrodes. The scale bar is 3 $\mu$m.  The mismatch between  the top and bottom electrodes are about 0.69 $\mu$m for the region A in the enlarged area while about 0.72 $\mu$m for the region B.	The normalized resistance $R(T)/R$(300 K)-1 v.s. $T$ with the current along $ab$ (b) and $c$ directions (c), respectively, under various magnetic fields $H$  parallel to the $c$ axis.
}
	\label{fig:Figure 2}
\end{figure}

Single crystals of WTe$_2$ were synthesized using a self-flux method. 
WTe$_2$  thin flake was transferred with the polyvinyl alcohol (PVA), based on our previous technique.\cite{zhang2021observation} Fig. 1  shows the schematics of the vertical contact fabrication and assembly process comprised of  the following steps:
(a) The electrodes are patterned when the mask is exposed to UV light.
(b) Gold is deposited to the SiO$_2$/Si substrate and the residual photoresist is removed with a standard lift-off process as depicted.
(c) PVA solution was spin-coated on the SiO$_2$/Si substrate with Au electrodes at a speed of 4000 rpm. It created a PVA film of about 900 nm. After 12 h in atmosphere, the PVA becomes dry and solid.  At room temperature, the Au electrodes are gently peeled off from the SiO$_2$/Si substrate by PVA\cite{cao2020movable,zhang2021observation}.
(d)  The WTe$_2$ thin flakes can be exfoliated using scotch tapes and attached to SiO$_2$/Si substrate prior to plasma cleaning.
(e) A piece of the Au/PVA strip is attached to PDMS and glass slide, under the microscope, they are aligned to the thin sample flake. After a short baking of 60$^o$C, the sample can be gently separated from the substrate and  grabbed by the PVA side.
(f) We align the Au electrodes part on the SiO$_2$/Si substrate with the WTe$_2$/Au/PVA/PDMS part on the glass slide under the microscope. After a short baking of 80$^o$C, the top PDMS is detached.
(g) After the PVA is washed away in deionized water and the part left is a Au/WTe$_2$/Au vertical structure on the SiO$_2$/Si substrate.
The optical microscope image of the vertical device is shown in the right part. Atomic force microscopy (AFM), Fig. 1(h), reveals the vertical channel region (in bright yellow colour) of the device. The width of the vertical channel is about 6 $\mu$m, the height profile in the inset shows that the thickness of the WTe$_2$ flake is 18.6 nm.

Fig. 2(a) shows the scanning electron micrograph (SEM) of the WTe$_2$ Van der Waals device.  The thin flake of WTe$_2$ is sandwiched between  Au electrodes. The top and bottom Au electrodes are laterally edge-to-edge separated by 3 $\mu$m from each other, respectively. The mismatch among top and bottom electrodes is inspected by the  magnified area in the inset. The mismatch length (d$_ s$) is about 0.69 $\mu$m and 0.72 $\mu$m, with a small inclination of about 4.3$\%$. The device can be used for measuring the planar and vertical transport properties at the same sample. For  the in-plane transport configuration, the electrodes 5 and 8 are used as current probes and inner electrodes 6 and 7 are connected to a voltmeter. While for out-of-plane transport configuration, the electrodes 2 and 6 are used as current probes and the electrodes 3 and 7 as voltage probes. The  resistivity was measured by two sets of lock-in amplifiers Stanford SR830  in a 14 T Oxford instrument Teslatron PT. To avoid the influence of noise from different directions, the out-of-plane transport measurements are performed subsequently after the in-plane measurements.





 \begin{figure*}
	\includegraphics[trim=0cm 0cm 0cm 0cm, clip=true, width=1\textwidth]{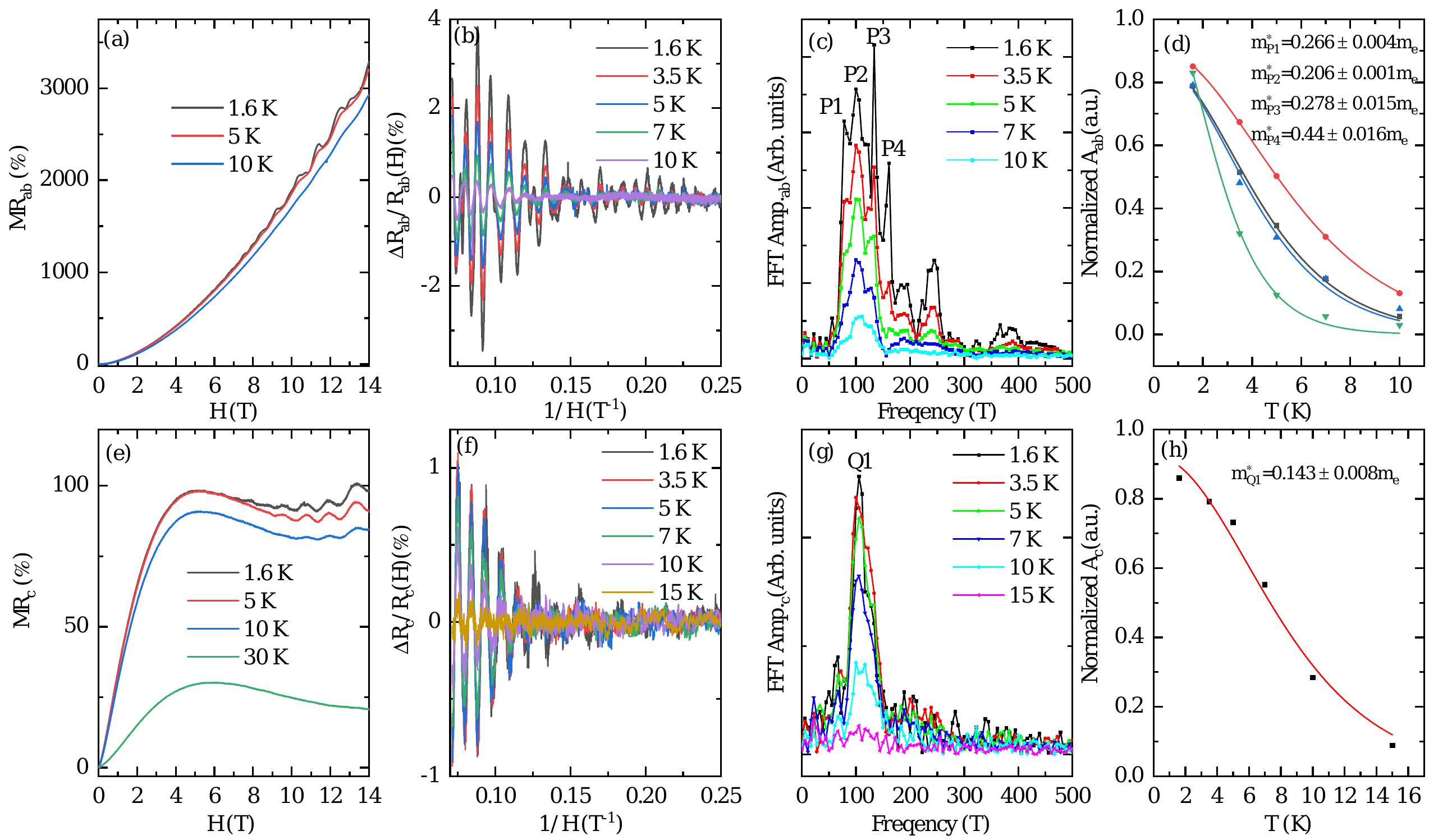}
	\caption{The magneto-transport behaviour in thin film WTe$_2$.
		The $H$ dependence of  the magnetoresistance  (MR) with the current along $ab$  (a) and $c$ directions (e) at different $T$.  SdH oscillations in  MR$_{ab}$ (b) and MR$_{c}$ (f).
	 The fast Fourier transform (FFT) spectrum of  MR$_{ab}$  (c) and MR$_{c}$ (g). The $T$ dependence of the normalized FFT amplitudes of MR$_{ab}$ (d) and MR$_{c}$  (h).
	}
	\label{fig:Figure 3}
\end{figure*}

\section{Results and discussions} \label{dis}

Fig. 2(b) and 2(c) show the temperature ($T$) dependence of  the normalized resistances ($R_{ab}$ and $R_{c}$) at the same WTe$_2$ sample under the external magnetic fields applied perpendicular to the plane and the applied current ($I$) along the in-plane and the out-of-plane directions, respectively. It is found that the $R_{ab}$  at 0 T decreases monotonically from room temperature as the temperature cools down, yielding a residual resistivity ratio (RRR) about 12.5 in Fig. 2(b).
The $R_{ab}$ shows a remarkable increase  after reaching a minimum at the turn-on temperature, when a magnetic field is applied. The turn-on temperature is found to be shifted to a higher temperature as increasing $H$ at  the investigated magnetic field range.  On the other hand, the RRR for $R_{c}$ at 0 T is  about 6.4 (Fig. 2(c)), which is less than the one for  $R_{ab}$. Furthermore, the $R_{c}$ show a negligible increase at low temperature at the investigated  $H\leq14$ T. It thereby suggests that the in-plane and  out-of-plane transport have distinct electronic properties.

Fig. 3(a) shows the $H$ dependence of the in-plane magnetoresistance (MR$_{ab}$) at various temperatures ($T$). The MR$_{ab}$ increases with $H$  without a sign of saturation, consistent with the literature.\cite{ali2014large,kong2015raman,zhu2015quantum,zhao2015anisotropic,xiang2015multiple}
At high $H$, the Shubnikov-de Haas quantum oscillations are observed in all the temperatures from 1.6 K to 10 K. The oscillatory component can be extracted by subtracting a second-order polynomial background over the MR.  The $\Delta R_{ab}/R_{ab}$(H) as a function of  $1/H$ is plotted in Fig. 3(b). Fast Fourier transformation (FFT) has been performed and the amplitude spectra are illustrated in Fig. 3(c). The frequencies of the thin flake show  four major peaks at 77.8, 100, 133, and 161 T, originated from  two pairs of electrons and hole pockets.\cite{zhu2015quantum,li2017evidence,cai2015drastic} The decrease and shift of the SdH oscillation frequencies are due to the spatial confinement contributing to the electronic structure in thin samples\cite{xiang2018thickness,li2017evidence}. In addition to the four distinct frequencies, a fifth frequency(P$_5$= 254 T) is detected. It is externally caused by the magnetic breakdown as suggested by a previous study.\cite{zhu2015quantum,cai2015drastic} The peaks correspond to the orthogonal cross-sectional area of the Fermi surface(FS) $A_F$, as described by the Onsager relation $F$=($\Phi_0$/2$\pi^2$)$A_F$, where $\Phi_0$ is the flux quantum. The effective cyclotron mass of carriers at the observed Fermi surface sheets is evaluated from the temperature dependence of normalized FFT amplitude using the  the Lifshitza-Kosevich (LK) formula.

$$ FFT  amp.\propto \frac{\alpha m^*T/H}{sinh(\alpha m^*T/H)}, $$
where $\alpha=2\pi^2K_B/e\hbar$ and $m^*=m/m_e$ is the effective mass. Normalized FFT amplitudes of the four peaks for the MR$_{ab}$ as a function of $T$ are plotted in Figs. 3(d). Their effective masses are estimated to be 0.266 $m_e$, 0.206 $m_e$, 0.278 $m_e$, 0.446 $m_e$ for oscillation peaks P$_1$, P$_2$, P$_3$, P$_4$, respectively.

When the current is applied through the vertical channel, we can probe the c-axis magnetoresistance transport properties in one single sample. The $H$ dependences of MR$_{c}$ under different $T$ are shown in Fig. 3(e). The MR$_{c}$ at 1.6 K increases with  $H$ and  saturates at large magnetic field, which is different with MR$_{ab}$ of WTe$_2$. The MR$_{c}$ shows oscillations  at the temperature range from 1.6 K to 15 K, but then disappear at higher temperatures. Similar to the MR$_{ab}$, the  oscillatory component of MR$_{c}$ is obtained after a background subtraction, and $\delta R_c/R_c$(H) is periodic with $1/H$ as shown in Fig. 3(f). One sharp peak Q$_1$ = 106 T is obtained for the MR$_{c}$ in Fig. 3(g). After a fitting with the normalized amplitude A$_c$, the effective mass is estimated to be 0.143 m$_e$ (Fig. 3(h)), smaller than  those obtained from the in-plane direction transport measurements. Since the oscillation peaks and effective mass are related to the Fermi surface, the Fermi surface obtained from in-plane transport is different from the out-of-plane transport. Thus the presence of different behaviours of MR$_{ab}$ is determined by the Fermi surface that can be  modified by the applied current directions.

In addition to the in-plane oscillation, we discover the quantum oscillation in the out-of-plane magnetoresistance.  One possible interpretation would be that the Fermi arcs on the top and the bottom surface contribute to a novel type of quantum oscillation phenomena.\cite{potter2014quantum}  However, the underlying origin needs further experimental evidence and theoretical analysis.
In bulk WTe$_2$, the resistivity  first increases with magnetic field at low magnetic field before it starts oscillating with suppressing out-of-plane magnetoresistance\cite{bi2018spin}. There is no previous report for out-of-plane magnetoresistance in WTe$_2$ thin films. The appearance of oscillation at a damping magnetoresistance in the ultrathin film is consistent with the bulk.


\section{Summary} \label{conc}
In summary, we introduced the fabrication of the vertically assembled Au/WTe$_2$/Au/device and characterize its out-of-plane and in-plane transport properties. A saturated out-of-plane magnetoresistance is observed at low temperatures while it remains unsaturated for the in-plane magneto-transport. There are quantum oscillations discovered in the out-of-plane transport in addition to the in-plane transport. A single peak in the out-of-plane direction oscillation is found. Our results suggest that the presence of different behaviours of magnetoresistance and the topology of Fermi surfaces of WTe$_2$ could be determined by applied current directions.


\begin{acknowledgments}
T.H. acknowledge the support of NSFC Grant No. 11574338. H.X. acknowledge the support of NSAF Grant No. U1530402.
Y.G. Shi acknowledge the support of  NSFC Grant  No. U2032204 and the Informatization Plan of Chinese Academy of Sciences(CAS-WX2021SF-0102).

\end{acknowledgments}

$^{*}$ liuys@baqis.ac.cn
$^{\dag}$ c.m.schneider@fz-juelich.de


\bibliography{WTe2}
\end{document}